\documentclass[preprint]{revtex4}
\usepackage{graphicx}
\usepackage{amssymb,amsmath}
\usepackage{verbatim}
\begin{document}

\title{Microscopic activity patterns in the Naming Game}
\author{Luca Dall'Asta}
\address{Laboratoire de Physique Th\'eorique (UMR du CNRS 8627),
B\^atiment 210, Universit\'e de Paris-Sud, 91405 ORSAY Cedex (France)}
\author{Andrea Baronchelli}
\address{Dipartimento di Fisica, Universit\`a ``La Sapienza'' and SMC-INFM,
P.le A. Moro 2, 00185 ROMA, (Italy)}
\email{luca.dallasta@th.u-psud.fr, andrea.baronchelli@roma1.infn.it}

\begin{abstract}
The models of statistical physics used to study collective phenomena in some interdisciplinary contexts, such as social dynamics and opinion spreading, do not consider the effects of the memory on individual decision processes. 
On the contrary, in the Naming Game, a recently proposed model of Language formation, each agent chooses a particular state, or opinion, by means of a memory-based negotiation process, during which a variable number of states is collected and kept in memory. 
In this perspective, the statistical features of the number of states collected by the agents becomes a relevant quantity to understand the dynamics of the model, and the influence of topological properties on memory-based models.   
By means of a master equation approach, we analyze the internal agent dynamics of Naming Game in populations embedded on networks, finding that it strongly depends on very general topological properties of the system (e.g. average and fluctuations of the degree). 
However, the influence of topological properties on the microscopic individual dynamics is a general phenomenon that 
should characterize all those social interactions that can be modeled by memory-based negotiation processes.   
\end{abstract}
\maketitle

\section{Introduction}\label{sec1}
Language Games are a class of simple models of population dynamics conceived to reproduce the processes involved in linguistic pattern formation inside a population of individuals~\cite{nowak1999,steels2000}. 
They have been profitably used in order to understand the origin and the evolution of language \cite{general}, and have found an important field of application in Artificial Intelligence, where the ultimate goal consists in modeling the self-organized collective learning processes in populations of artificial agents~\cite{steels1997,kirby2000}.
Recently, on the basis of these ingredients, a model called Naming Game has been put forward as a simple example of collective dynamics leading to the self-organized emergence of a communication system (i.e. linguistic conventions) in a population of interacting agents \cite{steels1995,baronka}.
The original definition of the model considers a population of agents that assign names to an object, trying to agree on a unique shared name by means of pairwise negotiations. 
The Naming Game may be applied to different contexts. For instance, it may be used to model the opinion spreading in a population of individuals that interact by means of negotiation, rather than imitation (as in the Voter model \cite{voter}). The concepts of memory and feedback on which the Naming Game is based are quite new in social dynamics and in statistical mechanics as well. They are at the origin of very interesting dynamical properties, some of them have motivated the present work. In particular, we will focus on the role of agents' memory, by means of which an agent can store several different states (or words, opinions, etc.) at the same time. The aim of this work is to provide a detailed statistical description of the internal dynamics of single agents in the Naming Game, studying their relation with the collective behavior of the model in its different dynamical regimes.

As many other models of social interaction, the Naming Game is a non-equilibrium model in which the system eventually reaches a stationary state. The dynamical evolution of these systems is usually characterized by a temporal region in which the system {\em reorganizes} itself followed by the sudden onset of a very fast {\em convergence} process induced by a symmetry breaking event. 
The Naming Game presents this type of dynamics when the agents are embedded in a mean-field like topology, i.e. a complete graph, and complex networks with small-world property, that are undoubtedly the most realistic cases for models of social interaction.
  
With respect to usual global quantities, studied in Refs.~\cite{baronka,NGnets,NGonSW}, the analysis of single agents activity  allows to investigate the connection between the learning process of the agents\footnote{We call "learning process" the dynamics of acquisition and deletion of states from the point of view of a single agent. The terminology is reminiscent of the original purposes of the Naming Game problem.} and the topological properties of the system. Interestingly, it turns out that, far from the convergence process, the shape of the distribution of the number of states stored by an agent, i.e. its memory size, depends on purely topological properties of the system (i.e. the first two moments $\langle k\rangle$ and $\langle k^2\rangle$ of the degree distribution). 
In particular, we show analytically, by means of a master equation approach, that homogeneous graphs yield exponential distributions, while heterogeneous networks, characterized by large fluctuations of the agents degree, give rise to half-normal distributions.

\begin{figure}
\centerline{
\includegraphics*[width=0.6\textwidth]{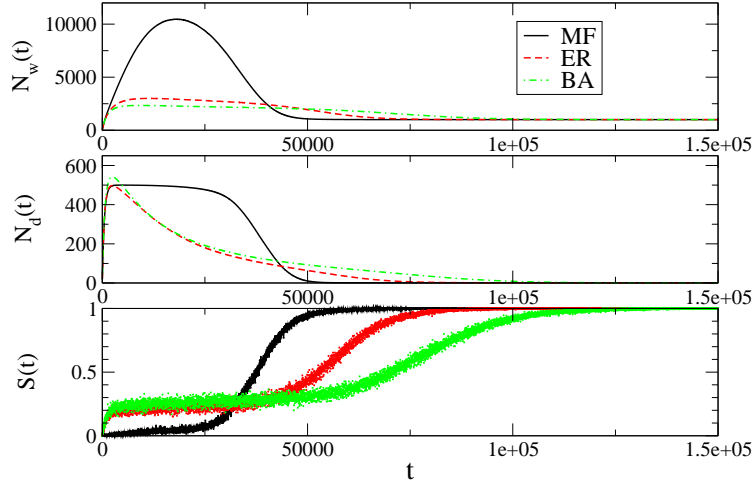}
}
\caption{Global behavior of the Naming Game on different topologies.   The complete graph (mean-field case) is compared to Erd\"os Renyi and Barab\`asi-Albert graphs, both with average degree $\langle k \rangle = 10$. In all cases, after an initial spreading of the different states, the dynamics goes through a period in which different states (whose total number is $N_d(t)$) are exchanged among the agents. Thus, the total number of states, $N_w(t)$, grows till a maximum and then start decreasing due to successful interactions which eventually lead the system to converge ($N_d(t)=1$, $N_w(t)=N$).
Finite connectivity allows for a faster initial growth in the success rate, $S(t)$. However, small-world properties give rise to the same exponential convergence observed in the fully connected graph. 
Data refers to populations of $N=1000$ agents.}
\label{figmodel}
\end{figure} 

During the convergence process, on the other hand, the master equation approach is still appropriate to describe agents internal dynamics,  but with qualitatively different results. All these systems tend to develop a power-law memory-size distribution, that is a signature of the convergence process, but it actually emerges only in the case of the complete graph (called `mean-field' case). In other topologies, the cut-off sets in too early for the power-law to be observed. 

Therefore, the analytical and numerical study of the memory-size distributions provides deep insights on the influence of the topology in the dynamics of the Naming Game. Moreover, the new findings are complementary to those already known from the analysis of global observables, and allow for a deeper understanding of the observed phenomena.
   
The paper is organized as follows. The next section is devoted to the description of the Naming Game model. Section~\ref{sec3} contains the main numerical results concerning the internal dynamics of individual agents in the Naming Game. In section~\ref{sec4}, the problem of determining agents internal dynamics is faced using a master equation approach.
Section~\ref{sec5} is devoted to illustrate in details some interesting cases. 
Conclusions on the relevance of the present work are reported in section~\ref{sec6}.

\section{The model}\label{sec2}

We consider the minimal model of Naming Game (NG) proposed in Ref.~\cite{baronka}.
A population of $N$ identical agents are placed on the vertices of a generic undirected network, while the edges identify the possible interactions between them.
An agent disposes of an internal inventory, in which it can store an a priori unlimited number of states.
As initial conditions we require all inventories to be empty. At each time step, a pair of neighboring agents is chosen randomly, one playing as ``speaker'', the other as ``hearer'', and negotiate according to the following rules:

\begin{itemize}
\item the speaker selects randomly one of its states and conveys it to the hearer;
\item if the hearer's inventory contains such a state, the two agents update their inventories so as to keep only the state involved in the interaction ({\em success});
\item otherwise, the hearer adds the state to those already stored in its inventory ({\em failure}).
\end{itemize}

The collective behavior of the system on different networks has been largely studied in Refs.~\cite{NGnets,NGonSW}. In particular, it turns out that essential quantities to describe the convergence process are the total number of states present in the system, $N_w(t)$, the number of different states, $N_d(t)$, and the success rate, $S(t)$, defined as the probability of a successful interaction at a given time. In Figure~\ref{figmodel} curves relative to complete graph (mean-field), Erd\"os-Renyi (ER) homogeneous random graph and Barab\`asi-Albert (BA) heterogeneous network are reported (see Refs.~\cite{mendes03,psvbook} for reviews of the networks models).
In the fully connected graph the process starts with an initial moderately fast (linear with time) spreading of states throughout the system followed by a longer period ($O(N^{1.5})$) in which states are exchanged among the agents. The total number of states then reaches a maximum and starts decreasing slowly till a point in which an super-exponential convergence leads the population to the adsorbing configuration in which all agents have the same unique state. 
On low-dimensional lattices and hierarchical structures, on the other hand, the model converges very slowly, and the reason is related to the formation of many different local clusters of agents with the same unique state, growing by means of coarsening dynamics \cite{NGletter}.
Finally, in the case of networks with finite average connectivity (sparse graphs), the initial dynamics is similar to that registered in low-dimensional regular structures, but the small-world property (i.e. average inter-vertex distance scaling as $\log N$ and presence of shortcuts connecting otherwise distant regions) boosts up the convergence process restoring the fast mean-field like cascade effect leading the system towards the global agreement. 

The present work, however, is addressed to study this model from a different and complementary point of view, focusing on the activity patterns of single agents. The next section is devoted to show some numerical results on the individual dynamics.

Before proceeding, a remark is in order. In heterogeneous networks, highly connected nodes (hubs) play a different role in the dynamics compared to low degree nodes. Indeed, as already pointed out for the Voter model~\cite{castellano}, the asymmetry of the NG interaction rules becomes relevant when  the degree distribution of the network, $p_k$, has long tails. When selecting the two interacting agents, the first node is thus chosen with probability $p_{k}$, while the hearer is chosen with probability $q_k = k p_k/\langle k\rangle$. Then the high-degree nodes are preferentially chosen as hearers, if the first extracted node is the speaker. We adopt this selection criterion, called {\em direct Naming Game}, since it fits realistic speaker-hearer interactions naturally. However other strategies are possible: one could first select the hearer and then the speaker ({\em reverse NG}), or more neutrally, an edge could be selected and the role of speaker and hearer assigned with equal probability among the two nodes ({\em neutral NG}). For further details on the consequences of pairs selection rules see \cite{NGnets}.  \\

\section{Numerical results on agents activity}\label{sec3}

In this section, we study numerically the activity of an agent focusing on the dynamics of its memory or inventory size, i.e. the number of states $n_{t}$ stored in the inventory of a node at the time $t$. 
In particular, the present analysis is conceived for populations on which we cannot clearly 
identify a coarsening process leading to the nucleation and growth of clusters containing quiescent agents (e.g. complete graph, homogeneous and heterogeneous random graphs, high-dimensional lattices, etc.) \cite{NGnets,NGonSW,NGletter}. 

\begin{figure}
\centerline{
\includegraphics*[width=0.6\textwidth]{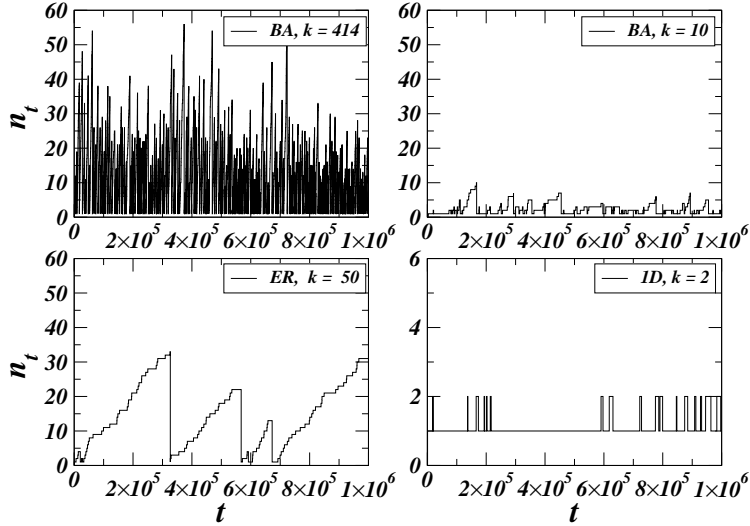}
}
\caption{Examples of temporal series of the number of states at a given node. Top) Series from a Barab\'asi-Albert (BA) network with $N=10^4$ nodes and average degree $\langle k\rangle =10$, for nodes of high degree (e.g. $k=414$) and low degree (e.g. $k=10$). Bottom) Series for nodes in Erd\"os-R\'enyi random graph ($N=10^4$, $\langle k \rangle = 50$) and in a one-dimensional ring ($k=2$).}
\label{fig1}
\end{figure}

Complex networks represent typical examples of such topological structures.
In other topologies, such as in low-dimensional lattices, the agents internal activity is limited by the small number of words locally available.  
An example of the different activity patterns in different topologies is reported in Fig.~\ref{fig1}.
Top panels show the different level of activity displayed by low and high degree nodes in a BA heterogeneous network.
The hubs are more active, being preferentially chosen as hearers, and they may reach larger inventory sizes (memory).
In homogeneous networks (bottom-left panel) all agents display approximately the same level of activity. In this case we reported an ER random graph with rather large average degree, so that the inventory may reach moderately large sizes. It is possible to verify with a magnification of the scales that the structure of the peaks is the same for all networks. The only topology displaying clearly different results is the regular one-dimensional lattice (bottom-right panel), in which the inventory size does not exceeds $2$ because of the coarsening process \cite{NGletter}.   

\begin{figure} 
\centerline{
\includegraphics*[width=0.6\textwidth]{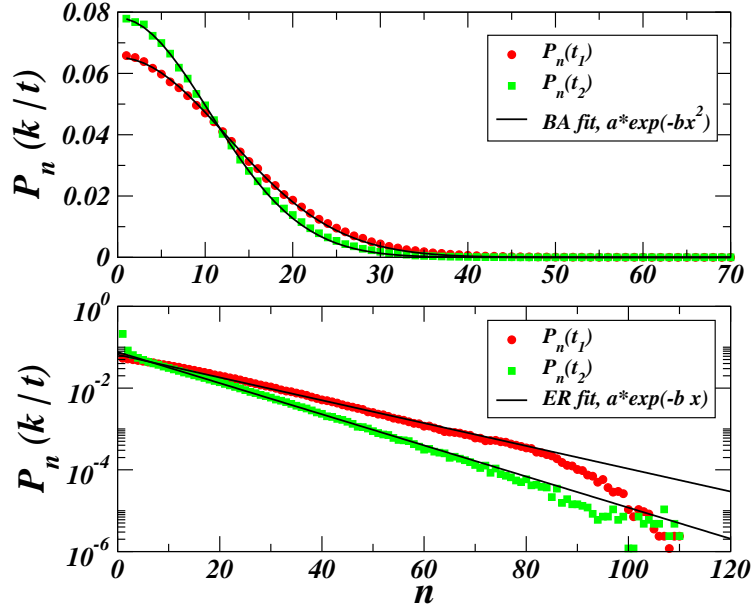}
}
\caption{Parametric dependence on time of the distribution of the number of states: the time has the effect of deforming the shape of the distributions, but does not change their functional description. Top) BA graph of $N=10^4$ nodes with $ \langle k \rangle = 10$. Only the set of nodes with $k > 150$ (hubs) is monitored. Histograms come from measurements at different times $t_1$ and $t_2$ with $t_2-t_1 = 5.10^5$ time-steps. Bottom) ER graph of $N=10^4$ nodes and $ \langle k \rangle = 10$. Measures refer to to the set of nodes with $k>70$. $t_2-t_1 = 4.10^5$ time-steps.}
\label{fig3}
\end{figure}

A quantity that clearly points out the statistical differences in the activity of the nodes depending on both their degree and the topological structure of the network is the probability distribution $\mathcal{P}_{n}(k|t)$ that a node of degree $k$ has a number $n$ of states in the inventory at the time $t$. 
The distribution is computed averaging over the class of nodes of given degree $k$ at a fixed time $t$. 
Fig.~\ref{fig3} displays typical inventory size distributions for the Naming Game on complex networks computed in the reorganization region that precedes the convergence.  
The top panel of Fig.~\ref{fig3} reports $\mathcal{P}_{n}(k|t)$ for the case of highly connected nodes in a heterogeneous network (the Barab\'asi-Albert network), whereas the bottom panel shows the same data for nodes of typical degree in a homogeneous network (the Erd\"os-R\'enyi random graph). From the comparison of the curves for different temporal steps (in the reorganization region), it turns out that in both cases the functional form of the distribution does not change considerably in time; the time $t$ enters in the distributions as a simple parameter governing their amplitude and the position of the cut-off. 

Moreover, in homogeneous networks the shape of the distribution does not actually depend on the degree of the node, since all nodes have degree approximately equal to the average degree $\langle k\rangle$. 
In the heterogeneous networks a deep difference exists between the behavior of low and high degree nodes. 
Low degree nodes have no room to reach high values of $n$, thus their distribution has a very rapid decay (data not shown); for high degree nodes, on the contrary, the distribution extends for more than one decade and its form is much clearer.

Apart from the behavior of low degree nodes, it is clear that the functional form of the distribution $\mathcal{P}_{n}(k|t)$ is different in homogeneous and heterogeneous networks. Homogeneous networks are characterized by  exponential distributions, while high degree nodes in heterogeneous networks present faster decaying distributions, that are well approximated by half-normal distributions (i.e. with Gauss-like shape).

Both cases of homogeneous and heterogeneous networks appear different from that of the mean-field model studied in Ref.~\cite{baronka}, in which the agents are placed on the vertices of a complete graph and, during the reorganization, the inventory size distribution is given by the superposition of an exponential and a delta function peaked around $n \sim \sqrt{N}$.  The reason of these differences will be elucidated in the next sections by means of an analytical approach to the problem.

\begin{figure} 
\centerline{
\includegraphics*[width=0.6\textwidth]{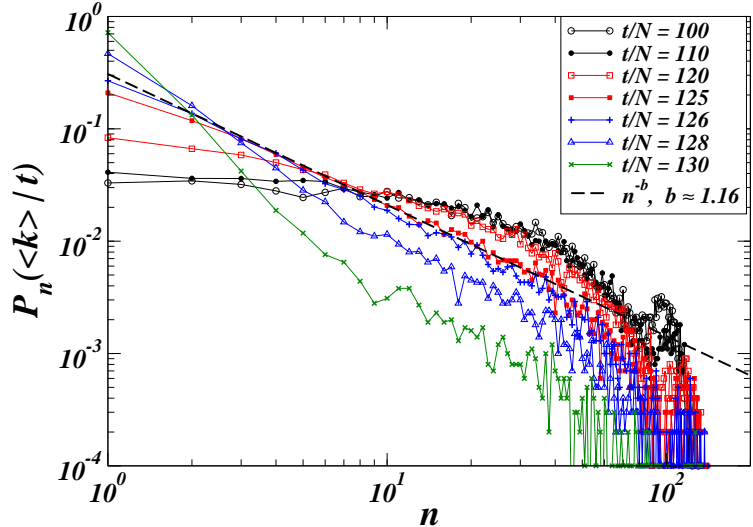}
}
\caption{Inventory size distribution for the Naming Game on the complete graph during the convergence process. At the beginning the peak at $n \sim \sqrt{N}$ gives way to a power-law, with exponent approximately $-1$, that rapidly becomes more and more steep at low values of $n$. The numerical data are obtained from a single run of the Naming Game on a complete graph of $N=10^4$ nodes, monitoring the whole temporal region of convergence. 
Note that we report single run experiments since the temporal fluctuations of the convergence process are rather large (see Ref.~\cite{baronka}), so that averaging over many runs may alter the real value of the power-law exponent.  
}
\label{fig5b}
\end{figure}

In contrast with the previous reorganization region, the main global quantities describing the dynamics accelerate when the system is close to the convergence: $N_{w}(t)$ converges to $N$, while $N_{d}(t)$ and $S(t)$ go to $1$, all with a super-exponentially fast process. 
Nevertheless, even in this region, the temporal scale of the global dynamics is much slower than that of agents activity, thus the fixed-time inventory size distribution $\mathcal{P}_{n}(k|t)$ is still a significant measure of the local activity.
In this case, the mean-field presents a more interesting phenomenology compared to sparse complex networks.
Fig.~\ref{fig5b} shows that, near the convergence, the complete graph develops a power-law inventory size distribution, with an exponential cut-off at $n \simeq \sqrt{N}$. Approaching the final consensus state the 
slope of the power-law becomes steeper and the cut-off moves backwards to $1$.\\
Similar power-law behaviors are not observed in any other topology even if it should be expected on homogeneous random graphs that, in the limit of large average connectivity, tend to the complete graph.
Numerical simulations instead show that, in the region of convergence, both homogeneous and heterogeneous complex networks (such as the ER model and the BA model) present an exponential distribution of the inventory size (data not shown). 

The numerical results reported in this section point out that the microscopic agents activity is closely related with the global dynamics and with the topological properties of the system. In the next section, we will show that, even if the dynamics of the number of states exhibited by a node is very complicated, mapping it on a jump process allows for some more rigorous results that give reason of the behaviors found in the numerical simulations.

\section{Master equation approach to agents internal dynamics}\label{sec4}

The jump process observed in the previous section and its statistics can be described using a master equation for the probability $\mathcal{P}_{n}(k,t)$ that an agent of degree $k$ has inventory size $n$ at time $t$. Formally, it reads 

\begin{eqnarray}
 \mathcal{P}_{n}(k,t+1)-\mathcal{P}_{n}(k,t)=& ~\mathcal{W}_{k}(n-1 \rightarrow n|t)\mathcal{P}_{n-1}(k,t)- \mathcal{W}_{k}(n\rightarrow n+1|t)\mathcal{P}_{n}(k,t)\\
\nonumber &~-\mathcal{W}_{k}(n\rightarrow 1|t)\mathcal{P}_{n}(k,t) ~~~~~ N_{d}(t) \geq n > 1\\ 
 \nonumber \mathcal{P}_{1}(k,t+1)-\mathcal{P}_{1}(k,t)=& ~\sum_{j=2}^{N_{d}(t)}\mathcal{W}_{k}(j \rightarrow 1|t)\mathcal{P}_{j}(k,t)- \mathcal{W}_{k}(1\rightarrow 2|t)\mathcal{P}_{1}(k,t)~, \label{rand}
\end{eqnarray} 

where $N_{d}(t)$ is the maximum number of different states present in the system at time $t$ and $\mathcal{P}_{n}(k,t)$
depends a priori explicitly on the time. Note that this equation describes the average temporal behavior of a class of  agents with the same degree $k$.\\
In order to get an expression for the transition rates, we call $C_{k}(t)$ the number of different words that are accessible to a node (of degree $k$) at time $t$, i.e. that are present in the neighborhood of the node. 
In the case of the complete graph, $C_{k}(t) = C(t) = N_{d}(t)$. 
The small-world property characterizing many complex networks ensures that the quantity $C_{k}(t)$ does not actually depend on $k$, since nodes with very different degree have access to the same set of different states (or words). Furthermore, the largest part of the states present in the system are accessible to all nodes. 
In small-world topologies, indeed, there is an initial spreading of words throughout the network that destroys local correlations. Consequently, we will safely approximate $C_{k}(t)$ with $C(t)$ and we can expect $C(t) \leq N_{d}(t)$ and proportional to it. 
The case of low-dimensional lattices is different since states can spread only locally, causing strong correlations between the inventories~\cite{NGletter}.

According to the numerical results exposed in section~\ref{sec3}, the behavior of $\mathcal{P}_{n}(k,t)$ allows to separate the evolution of the system in two regimes: a {\em reorganization region} extending from the maximum of $N_{w}(t)$ to the beginning of the convergence process, and a {\em convergence region}, involving the cascade process that leads the system to the final consensus state.
In addition, $\mathcal{P}_{n}(k,t)$ assumes different shapes for different topologies.

Interestingly, in both regions, the temporal dependence of the distribution turns out to be only parametric, i.e. it has the only effect of deforming the shape during the evolution. 
In other words, the actual distribution should be well approximated by a quasi-stationary solution $\mathcal{P}_{n}(k|t)$ of the master equation, only parametrically depending on the time. \\
This  means that the master equation can be solved by means of an {\em adiabatic approximation}, 
a method that is commonly used in the study of out-of-equilibrium systems with different time scales for the dynamics \cite{franz,ritort}.\\
In order to prove the validity of the adiabatic approximation, we need the expressions of the transition rates $\mathcal{W}_{k}(a\rightarrow b|t)$ from the inventory size $a$ to $b$ at time $t$, in both dynamical regimes and for different topologies.

\subsection{Transition rates in the reorganization region}\label{sec4a}

In a general context, the expressions of the transition rates can be derived from the probability of a successful  interaction, given by 

\begin{equation}\label{p_win_gen}
Prob\left\{ success \right\} = \frac{|S\cap H|}{n_{S}}, 
\end{equation}
\noindent where $|S\cap H|$ is the size of the intersection set between the inventories of the speaker and the hearer, and $n_{S}$ is the inventory size of the speaker.  
Note that expression in Eq.~\ref{p_win_gen} holds for every choice of the speaker-hearer pair, and its average over the population corresponds to the success rate $S(t)$. 
In the reorganization region, the intersection $|S\cap H|$ is on average close to zero and all states have approximately the same probability of appearing in the inventory of the speaker, justifying the assumption of uncorrelation of the inventories in all topologies with small-world property. 
From this assumption it turns out that the intersection is well expressed by $|S\cap H| \simeq n_{S} n_{H} / N_{d}(t)$ (where $n_{S}$ and $n_{H}$ are the inventory sizes of the speaker and the hearer). Indeed, the fraction of all accessible states that are present in the inventory of the speaker is $n_{S}/N_{d}(t)$; i.e. in each 
slot of the hearer's inventory there is a probability $n_{S}/N_{d}(t)$ of finding a given state. Since the average number of common states is given by the product of such probability and the hearer's inventory size $n_{H}$, the result for $|S \cap H|$ follows. 

The expressions of the transition rates are straightforward from the probability of a successful negotiation, $\frac{|S\cap H|}{n_{S}} \simeq n_{H}/N_{d}$.
Considering both the probabilities for the agent playing as hearer and speaker, the transition rate $\mathcal{W}^{r}_{k}(n\rightarrow 1|t)$ reads
\begin{equation}\label{pwinR}
\mathcal{W}^{r}_{k}(n\rightarrow 1|t) \simeq p_{k} \frac{\langle n \rangle_{t}}{C(t)} + q_{k} \frac{n}{C(t)}~,
\end{equation}
where the average inventory size $\langle n \rangle_{t}$ comes from the mean-field hypothesis for the neighboring sites of a node playing as speaker, that is actually correct in all small-world topologies, and $p_k$ and $q_k$ are the probabilities of playing as speaker and as hearer respectively. The index $r$ in ${W}^{r}_{k}$ is used to indicate that these transition rate are correct in the reorganization region.
The inventory size may increase only when the agent plays as hearer, i.e.  
\begin{equation}
\mathcal{W}^{r}_{k}(n\rightarrow n+1|t) \simeq q_{k} \left(1-\frac{n}{C(t)}\right)~.
\label{plooseR}
\end{equation} 

%
\begin{figure} 
\centerline{
\includegraphics*[width=0.6\textwidth]{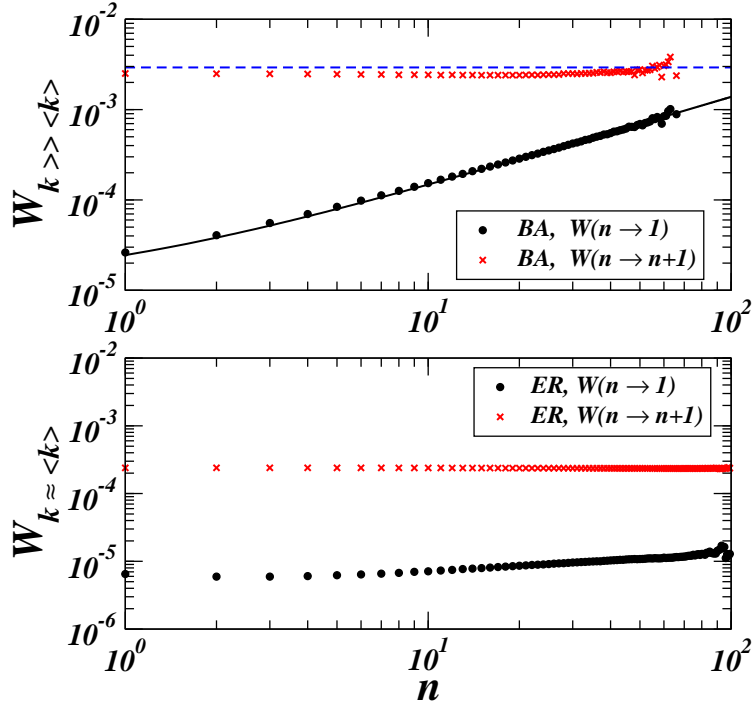}
}
\caption{Probability of winning and loosing (only the term causing an increase of the number of words) for BA and ER models.
Both with $N=5000$ nodes and $k\simeq 200$ (for a BA with $\langle k \rangle =10$) and $k\simeq 70$ (for a ER with $\langle k \rangle =50$). Data were obtained averaging over several runs ($3.10^4$) the probability of successful or unsuccessful interactions after $t=5.10^5$ time-steps from the beginning of the process. In fact, the time has also in this case only a parametric influence on the observed curves.
}
\label{fig6}
\end{figure}

In order to verify the above expressions for some specific cases, we have computed numerically the quantities $\mathcal{W}^{r}_{k}(n\rightarrow n+1|t)$ and $\mathcal{W}^{r}_{k}(n\rightarrow 1|t)$, in the case of a BA network of $N=5 \cdot 10^3$ nodes and $\langle k \rangle=10$ (top panel in Fig.~\ref{fig6}) and for an ER model with $N=5 \cdot 10^3$ nodes and $\langle k \rangle =50$ (bottom panel in Fig.~\ref{fig6}).\\
For heterogeneous networks, the numerical $\mathcal{W}^{r}_{k}(n\rightarrow 1|t)$ clearly show
a linear growth of the quantity with $n$, in agreement with Eq.~\ref{pwinR}, while the approximately constant behavior of  
$\mathcal{W}^{r}_{k}(n\rightarrow n+1|t)$ with $n$ can be fitted with an expression of the form Eq.~\ref{plooseR} only for very small values of $n/C(t)$. 
On the other hand, Fig.~\ref{fig6} (bottom) points out that in the case of homogeneous networks, in which all nodes have approximately the same behavior, both quantities are almost independent of $n$.
The different behaviors of the transition rates are responsible of the different shape of the probability distribution $\mathcal{P}_{n}(k|t)$. 

\subsection{Transition rates during the convergence process }\label{sec4b}

When the convergence process begins, the temporal behavior of all global quantities accelerates, and the expression of the success probability changes considerably. In all small-world topologies, the convergence is reached by means of a sort of cascade process, triggered by a symmetry breaking event in the space of the states (words, etc.) \cite{baronka}. 
The state involved in the symmetry breaking starts to win, becoming more and more popular among the inventories. 
At the end of the process, when the global consensus is reached, this is the only surviving state. 

According to this analysis, as the system is close to the convergence, most of the successful interactions involves the most popular state, while positive negotiations involving different states rapidly disappear. The statistical behavior of the quantity $\frac{|S\cap H|}{n_{S}}$ depends now only on the properties of the most popular word. The average size of the intersection set $|S\cap H|$ is well expressed by the probability $\alpha_{k}(t)$ of finding the most popular state (or word) in both the inventories. During the convergence process, $\alpha_{k}(t)$ is close to one. 
With this approximation we are neglecting the successful interactions due to less popular states, that we will show to have an effect for the dynamics on the complete graph (see section~\ref{sec5c}). \\
According with this argument, the transition rates assume the following form,    
\begin{eqnarray}\label{pwin_ploose1}
\mathcal{W}^{c}_{k}(n\rightarrow 1|t) \simeq p_{k} \frac{\alpha_{k}(t)}{n} + q_{k} \frac{\alpha_{k}(t)}{\langle n \rangle_{t}}~,\\
\label{pwin_ploose2} 
\mathcal{W}^{c}_{k}(n\rightarrow n+1|t) \simeq q_{k} \left(1-\frac{\alpha_{k}(t)}{\langle n\rangle_{t}}\right)~,
\end{eqnarray}
where the index $c$ is used to distinguish the expression of the transition rates during the convergence region from that of the reorganization regime. 

\subsection{Validation of the adiabatic approximation}\label{sec4c}

In both the reorganization and the convergence regions, the validity of the adiabatic approximation can be proved computing the characteristic relaxation time of the non-equilibrium process described by the master equation in Eq.~\ref{rand} with transition rates of Eqs.~\ref{pwinR}-\ref{plooseR} or Eqs.~\ref{pwin_ploose1}-\ref{pwin_ploose2}. 
Given the (continuous, for simplicity) master equation $\partial_{t} \mathcal{P}(t) = - \mathbf{W} \mathcal{P}(t)$, the relaxation time $\tau$ is defined as the inverse of the real part of the smallest non-zero eigenvalue $\lambda_{1}$ of the transition matrix $\mathbf{W}$.
The explicit diagonalization of the Markov transition matrix for a finite system may be demanding, but the order of magnitude of $\tau$ is easy to compute. We first note that $\mathbf{W} = p_{k} \bar{\mathbf{W}}$ in both cases.\\ 
In the reorganization region, when $C(t) \gg 1$, the real parts of the eigenvalues of $\bar{\mathbf{W}}$ are $\mathcal{O}(k/\langle k \rangle)$, thus $\lambda_{1} \propto q_{k}$, and the time necessary to reach the stationary state is $\tau \sim \mathcal{O}(1/q_{k})$. 
The argument holds even close to the consensus state, where $C(t)$, $\langle n\rangle_{t}$, and $\alpha_{k}(t)$ are of order $1$, since the smallest non-zero eigenvalue is still $\propto q_{k}$.  
Note that, in all complex networks $q_{k} > 1/N$, thus $\tau < N$. The time-dependent quantities involved in the expressions of the transition rates, such as $\langle n \rangle_{t}$ and $C(t)$ and $\alpha_{k}(t)$, vary on a slower timescale (the characteristic timescale of the global system is $t/N$), justifying the adiabatic approximation.

\subsection{General expression of the adiabatic solution in the two dynamical regions }\label{sec4d}

Mathematically, the adiabatic approximation consists in setting to zero the temporal derivative of the inventory size distribution, and looking at the stationary solution $\mathcal{P}_{n}(k|t)$, with parametric dependence on the time, that we call {\em adiabatic solution}. We compute the general adiabatic solution of the master equation in the two regions, while the most interesting cases are reported separately in the next section.

Let us first consider a general complex network in the reorganization region. 
Plugging the expressions of the transition rates $\mathcal{W}^{r}_{k}(n\rightarrow n+1|t)$ and $\mathcal{W}^{r}_{k}(n\rightarrow 1|t)$ into the stationary form of the master equation (Eq.~\ref{rand}), we get 
the following recursion relation,
\begin{equation}
\mathcal{P}_{n}(k|t) = \frac{q_{k} \left[ 1- \frac{n-1}{C(t)}\right]}{q_{k}\left[ 1- \frac{n}{C(t)}\right] + q_{k}\frac{n}{C(t)} + p_{k} \frac{\langle n\rangle_{t}}{C(t)}} \mathcal{P}_{n-1}(k|t)~.\label{rec_0} 
\end{equation}
Then, introducing $q_{k} = k p_{k}/\langle k\rangle$ $=$ $b(k)p_{k}$ and Eq.~\ref{rec_0} can be rewritten as
\begin{equation}
\mathcal{P}_{n}(k|t) = \frac{b(k) [1-\frac{n-1}{C(t)}]}{b(k)+\frac{\langle n\rangle_{t}}{C(t)}} \mathcal{P}_{n-1}(k|t)~. 
\end{equation}
Since $\frac{n-1}{C(t)} \ll 1$, we can write $1-\frac{n-1}{C(t)} \simeq e^{-\frac{n-1}{C(t)}}$, thus solving the recurrence relation,
\begin{equation}
\mathcal{P}_{n}(k|t) \simeq {s(k,t)}^{n-1} e^{-\frac{n(n-1)}{2C(t)}} \mathcal{P}_{1}(k|t)~,
\end{equation}
with $s(k,t) = b(k)/\left(b(k)+ \frac{\langle n\rangle_{t}}{C(t)}\right)$.
The normalization relation gives $\mathcal{P}_{1}(k|t)$. 
The controlling parameter of the curve is $s(k,t)$, that allows to tune the decay of the distribution between an exponential and a Gaussian-like tail. A change of variable $s(k,t)=1-\epsilon(k,t)$ (with $\epsilon(k,t) = \frac{ \langle n\rangle_{t}}{b(k) C(t)}$) makes evident that ${s(k,t)}^{n} \approx e^{- \epsilon(k,t) n}$, therefore the curve has the behavior
\begin{equation}
\mathcal{P}_{n}(k|t) \propto e^{-\epsilon(k,t) n -\frac{n(n-1)}{2C(t)}}~.
\label{apprexp}
\end{equation}
The linear term dominates when $\langle n \rangle_{t} \gg b(k)$, i.e. in homogeneous topologies, while the quadratic term governs the shape of the distribution for the high-degree nodes in heterogeneous networks ($\langle n \rangle_{t} \ll b(k)$).
This result is very interesting since it shows that heterogeneity is a necessary condition for agents to show a super-exponential decay in the inventory size distribution.

When we are in the convergence region, on the other hand, to get the form of the memory size distribution we must insert the Eqs.~\ref{pwin_ploose1}-\ref{pwin_ploose2} into the stationary version of Eq.~\ref{rand},
\begin{equation}\label{master2}
\begin{split}   
\frac{\partial\mathcal{P}_{n}(k|t)}{\partial t} = 0 = q_{k} \left(1-\frac{\alpha_{k}(t)}{\langle n\rangle_{t}}\right) \mathcal{P}_{n-1}(k|t) -  q_{k} \left(1-\frac{\alpha_{k}(t)}{\langle n\rangle_{t}}\right) \mathcal{P}_{n}(k|t) &\\
- \left[ p_{k} \frac{\alpha_{k}(t)}{n} + q_{k} \frac{\alpha_{k}(t)}{\langle n\rangle_{t}} \right] \mathcal{P}_{n}(k|t)~.&
\end{split}
\end{equation}
We get the following recursive relation,
\begin{equation}\label{cont_master}
 \mathcal{P}_{n}(k|t) = \left[ \frac{ 1 - \frac{\alpha_{k}(t)}{\langle n \rangle_{t}} }{1 + \frac{\alpha_{k}(t)}{b(k)}\frac{1}{n}}\right] \mathcal{P}_{n-1}(k|t)~,
\end{equation}
in which $b(k)= k/\langle k\rangle$.
The general solution is of the form
\begin{equation}\label{soluzConv}
\mathcal{P}_{n}(k|t) \propto {n^{-\frac{\alpha_{k}(t)}{b(k)}}}e^{- \frac{\alpha_{k}(t)}{\langle n \rangle_{t}} n}~,
\end{equation}
showing that near the convergence, the inventory size distribution may develop a power-law structure.
Nevertheless, in the section~\ref{sec3}, we stated that from numerical data there is no evidence of power-law behaviors on complex networks. This can be explained looking at the terms of Eq.~\ref{soluzConv}. In homogeneous networks, the power-law has exponent close to $1$ (since both $\alpha_{k}(t)$ and  $b(k)$ are of order $1$), but the cut-off imposed by the exponential distribution sets in at very low $n$, preventing the underlying power-law to be observed. The same argument holds for low-degree nodes in heterogeneous networks, but high-degree nodes should present sufficiently large inventories to see the power-law. However, in this case $b(k) \gg 1$, thus the exponent of the power-law is too small to be observed.\\
The only case in which we are able to observe a power-law inventory size distribution is that of the complete graph, that presents some peculiarities and will be discussed separately in the next section. 

\section{Adiabatic solution for some interesting cases}\label{sec5}

In this section, we study more in detail the effects of the topology on the adiabatic solution of the master equation making explicit calculations in three interesting cases: in the reorganization region, we consider the activity statistics of generic nodes in homogeneous random graphs and of hubs in  heterogeneous scale-free networks; in the convergence region, we focus on the purely mean-field behavior of agents placed on a complete graph.  

\subsection{The case of homogeneous networks}\label{sec5a}

As revealed by simulations reported in Fig.~\ref{fig6} (bottom) the transition rates for homogeneous networks in the reorganization region are almost independent of the number of states in the inventory. In homogeneous networks $q_{k} \simeq b(k) p_{k}$, with $b(k) \simeq \mathcal{O}(1)$, and the nodes are in general equivalent, thus the number of states is approximately the same for every node, i.e. $n \simeq \langle n \rangle_{t}$. 
The approximated expressions of the transition rates for a node of typical degree $k = \langle k\rangle$ are 
\begin{eqnarray}
\mathcal{W}^{r}_{k}(n\rightarrow 1|t)  \approx p_{k} \langle n \rangle_{t} (1+b(k)) /C(t) \approx 2 p_{k} \langle n \rangle_{t} /C(t)\\
\mathcal{W}^{r}_{k}(n\rightarrow n+1|t) \approx b(k) p_{k} \left(1- \frac{\langle n \rangle_{t}}{C(t)}\right) \approx p_{k}.\label{trans_hom}
\end{eqnarray}
Such  approximations are in agreement with the data reported in Fig.~\ref{fig6}-bottom.\\
The adiabatic condition for the master equation becomes
\begin{eqnarray}\nonumber
0 =& ~  \mathcal{P}_{n-1}(\langle k\rangle |t)- \mathcal{P}_{n}(\langle k\rangle|t)
- \langle n \rangle_{t} \frac{2}{C(t)} \mathcal{P}_{n}(\langle k\rangle|t) ~~~~~ n > 1\\ 
0 =& ~ \sum_{j=2}^{\infty} \frac{2}{C(t)} \langle n\rangle_{t}  \mathcal{P}_{j}(\langle k\rangle|t)-  \mathcal{P}_{1}(\langle k\rangle |t)~. \label{rand_hom}
\end{eqnarray} 
The solution by recursion is very simple,
\begin{equation}
\mathcal{P}_{n}(\langle k\rangle|t) \approx (1-\theta) {\theta}^{n-1}~, ~~~~~~~~~ \theta = \frac{1}{1+\frac{2\langle n \rangle_{t}}{C(t)}}~.
\label{statd_ER}
\end{equation}
Using the expansion of logarithm $\log(1-\epsilon) \simeq -\epsilon$, with $\epsilon = 1-\theta \simeq 2 \langle n\rangle_{t} /C(t)$, the previous formula gives the following exponential decay for the distribution of the number of states,
\begin{equation}
\mathcal{P}_{n}(\langle k \rangle|t) \simeq \frac{2 \langle n\rangle_{t}}{C(t)} e^{-\frac{2\langle n\rangle_{t}}{C(t)} n}~. 
\label{distrER}
\end{equation}
The exponential decay is in agreement with the numerical data. 
Knowing the complete form of the distribution (i.e. with the correct normalization prefactor), we can also roughly estimate $\langle n \rangle_{t}$ and $C(t)$, at fixed time $t$, from a self-consistent relation for $\langle n\rangle_{t}$, 
From Eq.~\ref{distrER}, we compute the approximate average value of $\langle n\rangle_{t}$, i.e.
\begin{equation}
\langle n \rangle_{t} \approx \int_{1}^{\infty} n \mathcal{P}_{n}(\langle k \rangle|t) dn~,
\end{equation}
and we get the self-consistent expression
\begin{equation}\label{self_n}
\langle n \rangle_{t} \simeq \left(\frac{C(t)}{2\langle n\rangle_{t}}\right) \left(1+\frac{2 \langle n\rangle_{t}}{C(t)}\right) e^{-\frac{2 \langle n \rangle_{t}}{C(t)}}~. 
\end{equation}
Now, introducing in Eq.~\ref{self_n} the numerical value of $\langle n\rangle_{t}/ C(t)$, it is possible to verify that the orders of magnitude of both $\langle n \rangle_{t} \sim \mathcal{O}(10)$ and $C(t) \sim \mathcal{O}(10^2)$ are in agreement with their numerical estimates.

\subsection{High-degree nodes in heterogeneous networks}\label{sec5b}

Now we pass to describe the dynamics of the hubs in heterogeneous networks in the reorganization region of the system. In a direct Naming Game, a hub is preferentially chosen as hearer, by a factor $b(k) = k/\langle k \rangle \gg 1$, then in the transition rates we can neglect the terms associated with the speaker. 
We consider the following approximated expressions
\begin{eqnarray}
\mathcal{W}^{r}_{k}(n\rightarrow 1|t) \simeq  q_{k} \frac{n}{C(t)}~, \label{phubsBA1}\\
\mathcal{W}^{r}_{k}(n\rightarrow n+1|t) \simeq  q_{k} \left(1- \frac{n}{C(t)}\right) \simeq q_{k}~, \label{phubsBA2}
\end{eqnarray}
in which the last approximation is justified by the fact that, in general, $n/C(t) \ll 1$. 
Inserting realistic values of $q_{k}$ and $C(t)$, the Eqs.~\ref{phubsBA1}-\ref{phubsBA2} are in agreement with the behaviors coming from the fit of the corresponding curve in Fig.~\ref{fig6} (top). \\
We can easily compute the adiabatic solution $\mathcal{P}_{n}(k|t)$ from Eq.~\ref{rand}
\begin{equation}
0= q_{k} \mathcal{P}_{n-1}(k|t) - \left(q_{k} + q_{k} \frac{n}{C(t)} \right)\mathcal{P}_{n}(k|t)~,
\end{equation}
and we find recursively
\begin{eqnarray}
\mathcal{P}_{n}(k|t) &= \frac{C(t)}{C(t)+n} \mathcal{P}_{n-1|t}(k) = \frac{{C(t)}^2}{(C(t)+n)(C(t)+n-1)} \mathcal{P}_{n-2}(k|t) \\
\quad & \approx \frac{{C(t)}^{n-1}\Gamma(C(t)+2)}{\Gamma(C(t)+n+1)} \mathcal{P}_{1}(k|t)~.
\end{eqnarray}
Now, from the closure relation $\sum_{n=1}^{\infty}\mathcal{P}_{n}(k|t) = 1$ we get the expression of $\mathcal{P}_{1}(k|t)$, and the final form for  $\mathcal{P}_{n}(k|t)$ becomes
\begin{equation}
\mathcal{P}_{n}(k|t) = \frac{{C(t)}^{n-1}}{\Gamma(C(t)+n+1)} {C(t)}^{C(t)+1} e^{-C(t)}\left[ \frac{\Gamma(C(t)+1)}{\gamma(C(t)+1,C(t))}\right],
\label{statd_BA}
\end{equation}
where $\gamma(a,x)$ is the lower incomplete Gamma function. 
The functional form of the stationary distribution is complicated, but exploiting Stirling approximations for Gamma functions we can easily write it into a much simpler form.
Indeed, using the expression $\Gamma(x) \approx \sqrt{2\pi} e^{-x} x^{x-1/2}$ and the representation via Kummer hypergeometric functions for the incomplete Gamma function $\gamma(a,x)$, we find that
\begin{equation}
\lim_{x\rightarrow +\infty} \frac{\Gamma(x+1)}{\gamma(x+1,x)} = const \simeq 2~,
\end{equation}
and this value is correct in the range of $x=C(t) \gg 1$. Finally, using the asymptotic series expansion of $\Gamma(x+n+1)$ for large $x$, we get an expression that can be formally written as
\begin{equation}
  \Gamma(x+n+1) \approx \sqrt{2\pi} e^{-x} x^{x+n+1/2} 
 \times \left\{\mathcal{O}(1) + Q[\mathcal{O}((n+1)^2)] x^{-1} + Q[\mathcal{O}((n+1)^4)] x^{-2} + \dots \right\}~, 
\end{equation}
in which  $Q[\mathcal{O}((n+1)^l)]$ is a polynomial in $(n+1)$ of maximum degree $l$. Now, we can do the resummation of the series keeping at each order $k$ in $x$ only the highest term in the polynomial in $(n+1)$, whose coefficient is $2^{-k}/k!$,
\begin{equation}
 \Gamma(x+n+1) \approx  \sqrt{2\pi} e^{-x} x^{x+n+1/2} \sum_{k=0}^{\infty}\frac{x^{-k}{(n+1)}^{2k}}{k! 2^{k}} = \sqrt{2\pi} e^{-x +{(n+1)}^{2}/2x} x^{x+n+1/2}~.    
\end{equation}
Putting together all the ingredients, we find that a good approximation of the distribution of the number of words is given by (the half-Normal distribution)
\begin{equation}
\mathcal{P}_{n}(k|t) \simeq \sqrt{\frac{2}{\pi C(t)}} e^{\frac{-{(n+1)}^{2}}{2C(t)}}~.
\label{distrBA} 
\end{equation}
Fitting numerical results in Fig.~\ref{fig6} (top) with this expression provides values for $C(t) \sim \mathcal{O}(10^2)$, showing that, as expected, on the BA model $C(t) < N_{d}(t) \sim \mathcal{O}(10^2 \div 10^3)$. 

\subsection{Power-laws on the complete graph}\label{sec5c}

The last interesting case consists in studying the inventory size distribution for agents on the complete graph.
In the reorganization region, the mean-field dynamics is characterized by a large fraction of agents with $\mathcal{O}(\sqrt{N})$ states in their inventories and another smaller fraction with exponentially distributed inventory sizes. 
The existence of a peak at $\mathcal{O}(\sqrt{N})$ comes from the initial accumulation process 
(see Ref.~\cite{baronka}), while the exponential part of the distribution is produced during the following reorganization regime. Since the most of the agents have $\mathcal{O}(\sqrt{N})$ states and the intersection between inventories is close to zero, we can write the following transition rates 
\begin{eqnarray}
\mathcal{W}^{r}_{k}(n\rightarrow 1|t)  \approx \frac{2}{N} \frac{1}{\sqrt{N}}\\
\mathcal{W}^{r}_{k}(n\rightarrow n+1|t) \approx \frac{1}{N} \left(1- \frac{1}{\sqrt{N}}\right) \label{trans_complete}
\end{eqnarray}
With the usual recurrence relation we compute the following adiabatic solution,
\begin{equation}\label{mean-field1}
\mathcal{P}_{n}(k|t) \propto f(t)\delta(n-\sqrt{N}) + \left(1-f(t)\right) e^{-\frac{2}{\sqrt{N}}n}~,
\end{equation}
with $f(t)$ is the fraction of agents around $\sqrt{N}$ that tends to zero the convergence

The interesting region is however the last one, during the convergence process, in which the inventory size distribution of the mean-field system develops a power-law structure.
In Eq.~\ref{soluzConv}, we have shown that the expected distribution in the convergence region presents a power-law, that in the particular case of the complete graph should have an exponent close to $1$ (since $\alpha_{k}(t) \simeq 1$). 
Nonetheless, Fig.~\ref{fig5b} reveals that the slope $-1$ is correct only at the beginning of the convergence process, while later the slope seems to increase, developing a bump in the range of small inventory sizes.
Starting from the previous remark on the mixed distribution emerging in the reorganization region, we explain how the alteration of the power-law is due to the superposition of an exponential distribution. \\
During the convergence process, the agents having access to the most popular state behaves following the transition rates in Eqs.~\ref{pwin_ploose1}-\ref{pwin_ploose2} and their activity is at the origin of the power-law in $\mathcal{P}_{n}(k|t)$. The other agents, that have no access to the most popular state maintain an inventory of size about $\sqrt{N}$ and fall to $1$ if they get a successful interaction. In other words, they keep on playing as in the reorganization region, generating an exponential distribution of the inventory sizes. Even if the fraction of these agents decreases in time, the superposition of the exponential on the power-law has the immediate effect of increasing the slope of the power-law at low $n$.\\
In summary, we have provided an explanation of the behavior of the activity patterns of the Naming Game on the complete graph, pointing out some fundamental differences with respect to generic complex networks. 
  
\begin{figure} 
\centerline{
\begin{tabular}{|c|}\hline \\ \includegraphics[width=0.45\textwidth]{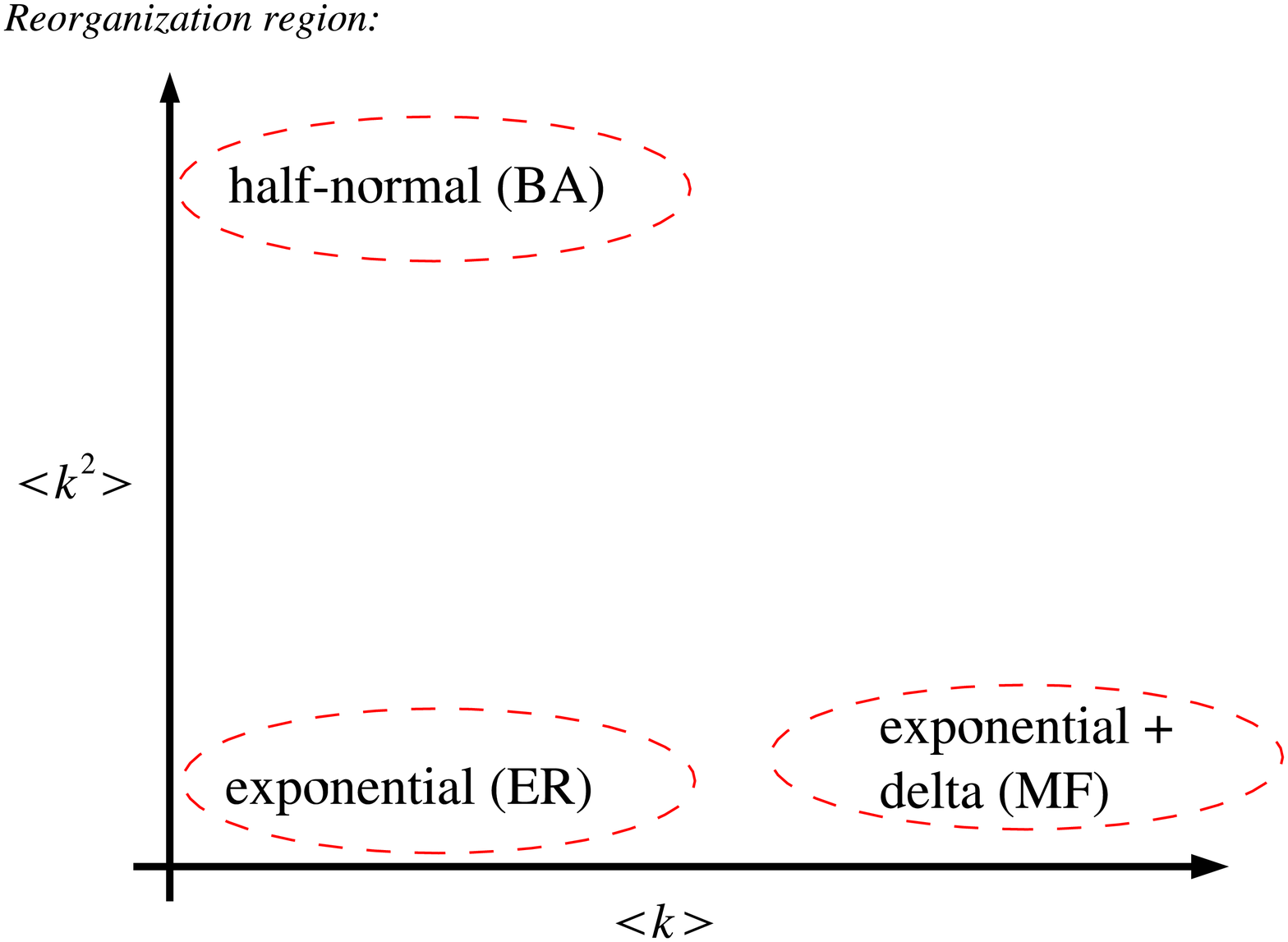}\\ \hline\end{tabular}~~
\begin{tabular}{|c|}\hline \\ \includegraphics[width=0.45\textwidth]{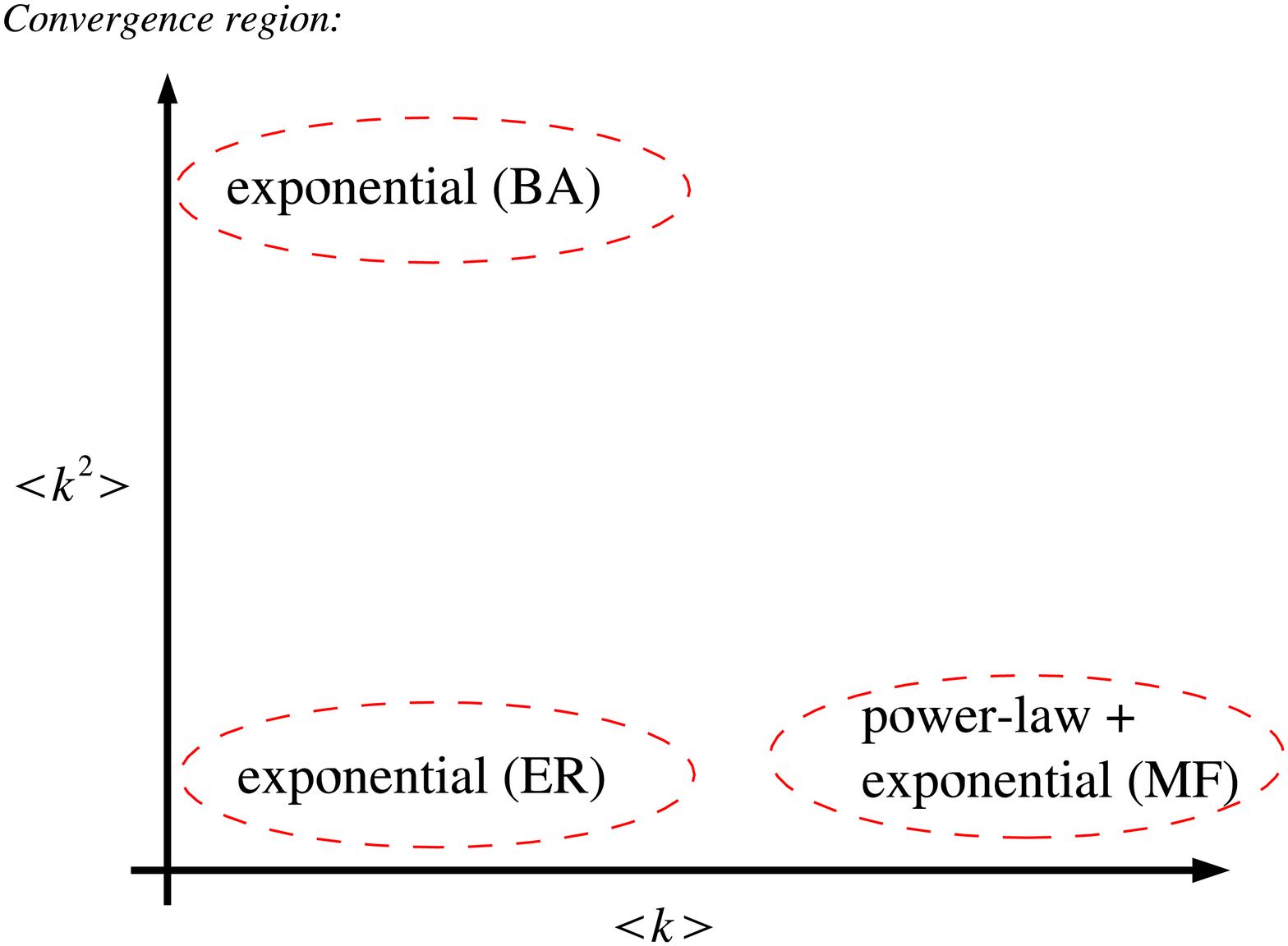}\\ \hline\end{tabular}
}
\caption{Phase plane like pictures in which the topological affects on the microscopic activity of the Naming Game are summarized. Left figure displays the situation in the reorganization region, in which the major effect is due to the increase of the degree fluctuations (the memory size distribution passes from an exponential to a half-normal distribution. In the right panel, we show the same picture for the convergence region, in which the final cascade process of convergence porduces a power-law like memory size distribution. Such a distribution is however visible only in the purely mean-field case, while on generic complex networks is covered up by exponential terms.
The region at both large average degree and fluctuations is difficult to be explored, but should correspond to mixed distributions in which all previously classified behaviors may be observed. 
}
\label{phase_plane}
\end{figure}

\section{Conclusions}\label{sec6}

We have studied the microscopic activity patterns in a population of agents playing the Naming Game proposed in Ref.~\cite{baronka}.
Previous work pointed out that the non-equilibrium dynamical behavior of the model presents very different features depending on the underlying topological properties of the system\cite{baronka,NGnets,NGonSW}. The analysis, however, were focused on the behavior of global quantities, while in the present work we have investigated the microscopic activity patterns of single agents. 
Indeed, by means of numerical simulations and analytical approaches, we have shown that the negotiation process between agents is at the origin of a very rich internal activity in terms of variations of the inventory size. 
More precisely, our analysis has focused on the instantaneous activity statistics described by the distribution $\mathcal{P}_{n}(k|t)$ that an agent of degree $k$ has an inventory of size $n$ at time $t$. We have been able to explain its behavior in function of both the global temporal evolution and the underlying topology of the system.

Apart from an initial transient, the dynamics of the Naming Game can be split in two temporal regions, namely the reorganization part and the convergence part. 
Fig.~\ref{phase_plane} summarizes our findings, showing the microscopic activity statistics in function of the first two moments of the degree distribution $P(k)$, which turn out to be essential features of complex networks affecting the dynamics of the $\mathcal{P}_{n}(k|t)$.
In the left panel of Fig.~\ref{phase_plane} we sketch the relation between topology and single agent activity in the reorganization region. Increasing the heterogeneity of the nodes the $\mathcal{P}_{n}(k|t)$ shifts from an exponential to a super-exponential (half normal) regime. Increasing $\langle k \rangle$ while preserving the homogeneity of the nodes, on the other hand, leads to a superposition of an exponential and a delta at $\sqrt{N}$. A class of distributions mixing up all these features is observed for networks with diverging average degree and fluctuations (top-right corner of the plane). 
A similar summary describes the effect of the topology in the convergence region (Fig.~\ref{phase_plane}, right panel): increasing the average degree, the distribution moves from exponential to a superposition of an exponential and a power-law, while larger fluctuations destroy the power-law leaving only an exponential distribution. 

In general, the influence of topological properties of complex networks on the dynamical properties of processes taking place on them is the object of a vast interest in statistical physics community. However, only global properties are usually considered. Here, we have focused on the internal dynamics of single agents, and we have found results providing explanation for the strong converging property of the corresponding global dynamics.
Indeed, one of the most interesting aspects of the Naming Game is exactly that the number of states an agent can store is not fixed a priori and the update rule involves a memory-based negotiation process. This is a relevant difference with most of the well known models in various fields of statistical mechanics or opinion dynamics, such as the voter or the Axelrod models, and we have investigated its deep consequences on the global behavior of the system.

A last remark concerns the comparison with usual statistical mechanics models. In this regard, it is useful to shift our perspective and look at the waiting time between successive decision events. In the present case, a decision event corresponds to a successful interaction, so that the waiting time is directly proportional to the inventory size. In the non-equilibrium glauber dynamics, for instance, a decision event is commonly associated to a spin flip. 
The corresponding waiting time is exponentially distributed during the dynamics (poissonian dynamics), but close to the convergence (to the ferromagnetic state) the waiting time between two flips may diverge, and its distribution assumes a power-law shape. 
As we have shown, a similar behavior is observed and proved for the inventory size distribution in the mean-field Naming Game. 
The inventory size statistics in the Naming Game can be thus compared to waiting time statistics in other models.  According to our analysis, it should be interesting to further investigate the relation between topology and individual waiting time statistics in other models of collective dynamics presenting similar non-poissonian individual activity. 

\section*{Acknowledgments} 
The authors thank A. Barrat and V. Loreto for many
useful discussions.  L.D. is partially supported by the EU under contract
001907 (DELIS). A.B. is partially
supported by the EU under contract IST-1940 (ECAgents).

\end{document}